\documentclass[12pt]{article}

\def\ii{\'{\i }}
\usepackage{graphics}
\usepackage{epsfig}
\usepackage{fancyhdr}
\usepackage{amssymb}
\usepackage{amsmath}
\begin{document}
\title{Evolution of particle density in high-energy pp collisions}
\author{ I. Bautista $^{\dag *}$ , C. Pajares \footnote{IGFAE
and Departamento de F\ii sica de Part\ii culas, Univ. of Santiago
de Compostela, 15782, Santiago de Compostela, Spain} and  J. Dias de Deus
\footnote{CENTRA, Departamento de F\ii sica, IST, Av. Rovisco
Pais, 1049-001 Lisboa, Portugal} }

 \maketitle

\begin{abstract}
We study the evolution of the particle density, $dn/d\eta$ at
fixed $\eta$, with the beam rapidity $Y$ in the framework of string percolation model.
Our main results are: $(i)$ The width of the "plateau" increases
proportionally to $Y$, $(ii)$ limiting fragmentation is
violated, and $(iii)$ the particle density, reduces to a
step function.

\end{abstract}

 We work in the framework of the string percolation and one of the main results is the presence of an extended "plateau" with a
  length $\Delta y$ proportional to $Y (\Delta y \simeq 1.4 Y)$, where $Y$ is the beam
  rapidity in the center of mass system: $Y\equiv ln(\sqrt(s)/m_{p})$. This result
  favors saturation models [1-4] with formation of longitudinal fields
  (flux tubes, effective strings) and naturally explains
 the presence of long range (pseudo-rapidity) correlations and the ridge phenomenon.
  See [5] for a discussion. Consequences of dominance of the "plateau" are the violation of limiting fragmentation and evolution towards a step function. 

In the string percolation model multi-particle production in a two nuclei collision
 at high energy can be explained in terms of the produced
strings along the collision axis, between the projectile and target.
These strings decay into new ones by $q\bar{q}$ or $qq$- $\bar{q}\bar{q}$ pair
production and subsequently hadronize to produce the observed hadrons. Due
to confinement, the color of these strings is confined to a small area in
transverse space $S_{1} = \pi r_{0}^{2}$ , with $r_{0} \simeq 0.2$- $0.3 $ fm.
 With increasing energy and/or atomic number of the colliding particles, the number
  of strings $N_{s}$ grows and they start to overlap forming clusters, very much
   like disks in two-dimensional percolation theory. At a certain critical density,
    a macroscopic cluster appears, which marks the percolation phase transition [6] [7].

In string percolation the relevant quantity is the transverse impact parameter density
$\eta^{t}$ which, in the case of pp
 collisions, we write as
\begin{equation}
\eta^{t}\equiv(\frac{r_{0}}{R_{p}})^{2}\bar{N}_{s},
\end{equation}
where $r_{0}$ is the single string transverse size, $R_{p}\simeq
1$ fm is the proton transverse size and $\bar{N}^{s}$ is the
average number of single strings. For values of $\eta^{t}$ below
the critical 2- dimensional density for percolation,
$\eta^{t}_{c}\simeq 1.15- 1.5$ [8], the formed strings do not
interact and collective effects are not present. For values
$\eta^{t}\gtrsim \eta^{t}_{c}$ one observes the formation
 of long strings due to fusion, stretching between the beam and the target.
 We assume that $\eta^{t}\gtrsim \eta^{t}_{c}$.

The particle density $dn/dy$ at mid-rapidity is related to the average
number $\bar{N}_{s}$ of strings
\begin{equation}
\frac{dn}{dy}\sim F(\eta^{t})\bar{N}_{s},
\end{equation}
where $F(\eta^{t})$ is the color reduction factor [9], due to
color summation of random colors,
\begin{equation}
F(\eta^{t})\equiv \sqrt{\frac{1-e^{-\eta^{t}}}{\eta^{t}}}.
\end{equation}
For small $\eta^{t}$, $F(\eta^{t})\rightarrow 1$, and for
$\eta^{t}\gtrsim \eta^{t}_ {c}$, $F(\eta^{t})\rightarrow
\frac{1}{\sqrt{\eta^{t}}}$. Starting with an exponential growth of the average
number $\bar{N}_{s}$ of strings,
\begin{equation}
\bar{N}_{s}\sim e^{2 \lambda Y},
\end{equation}
with $\lambda \simeq 0.2-0.3$ [10,11], we write for the particle
density, at $y\simeq 0$,
\begin{equation}
\frac{dn}{dy}\sim e^{\lambda Y},
\end{equation}
and for the full rapidity distribution, [12,13],
\begin{equation}
\frac{dn}{dy}|_{pp}=a e^{\lambda Y} \frac{dn}{dy}|^{s},
\end{equation}
where $\frac{dn}{dy}|^{s}$ is the single string density, $\frac{dn}{d \eta}|_{pp}$
 is the pp density, and $a$ is a constant, depending on the nature of produced particles.

Note that the exponential behavior in (4), $\bar{N}^{s}\sim e^{2 \lambda Y}$, is not an assumption but results from a simple application of conservation of energy to $dn/dy$ - making use of (2), (3) and the high density limit - to obtain (4) with $\lambda \simeq 2/7$, [11]. The rise of the multiplicity plateau is not proportional to $Y$ but to $\exp(\lambda Y)$.

Following [12,13] we write, for $\eta \geq 0$,
\begin{equation}
\frac{dn}{dy}|^{s}=\frac{1}{e^{\frac{\eta-(1-\alpha)Y}{\delta}}+1},
\end{equation}
$\alpha$ and $\delta$ being free parameters. Finally, using the Jacobean $J$ of the
 $y \rightarrow \eta$  transformation, we construct $dn/d\eta |_{pp}$,
\begin{equation}
\frac{dn}{d\eta}|_{pp}=J\frac{dn}{dy}|_{pp},
\end{equation}
with $J=\frac{cosh \eta}{\sqrt{k+sinh^{2} \eta}}$,
$k=\frac{m^{2}+p_{T}^{2}}{p_{T}^{2}}$. and by assumption fix $k$
at the effective value 1.2. It corresponds to $m_{\pi}\simeq 0.14
$ GeV and $\bar{p}_{\pi}\simeq 0.3$ GeV also in agreement with
[14].

In Fig. 1 we show our fits to $dn_{ch}/d\eta$ for pp collisions
(excluding single diffraction) from 53 GeV to 1.8 TeV [15], and at
LHC energies [16]. The free parameters where fixed at
$\lambda=0.23 \pm 0.005 $, $\delta=0.61 \pm 0.15$, $\alpha=0.27
\pm 0.03$, $a=0.8 \pm 0.2$.
\begin{figure}
\begin{center}
      \resizebox{120mm}{!}{\includegraphics{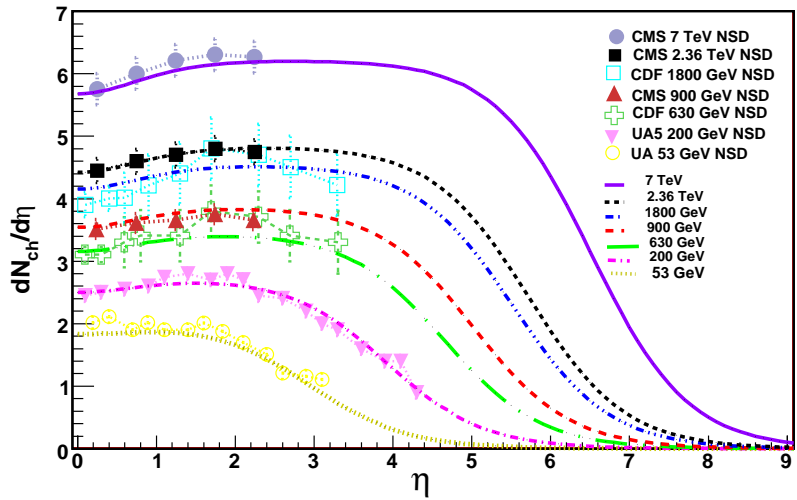}}
    \caption{$dN_{ch}/d\eta$ at different $\sqrt{s}$. For data points see references
     [15],[16]. Data from UA5, 546 GeV and UA1 at 540 GeV being
 inconsistent, were not included. Data from P238, 630 GeV are also not included.
 Note that the $\sqrt{s}=$ 2.36 TeV and 0.9 TeV plotted is from CMS for clarity
 of the plot the ALICE results also follow in these range and they differ form
 the CMS data for less than 0.2. }
    \label{test4}
\end{center}
\end{figure}

In Fig. 2 we show the evolution with $Y$ at different values of
pseudo-rapidity, $\eta$. Evolution within the "plateau" is slower than evolution
in the "fragmentation" region.
\begin{figure}
\begin{center}
      \resizebox{120mm}{!}{\includegraphics{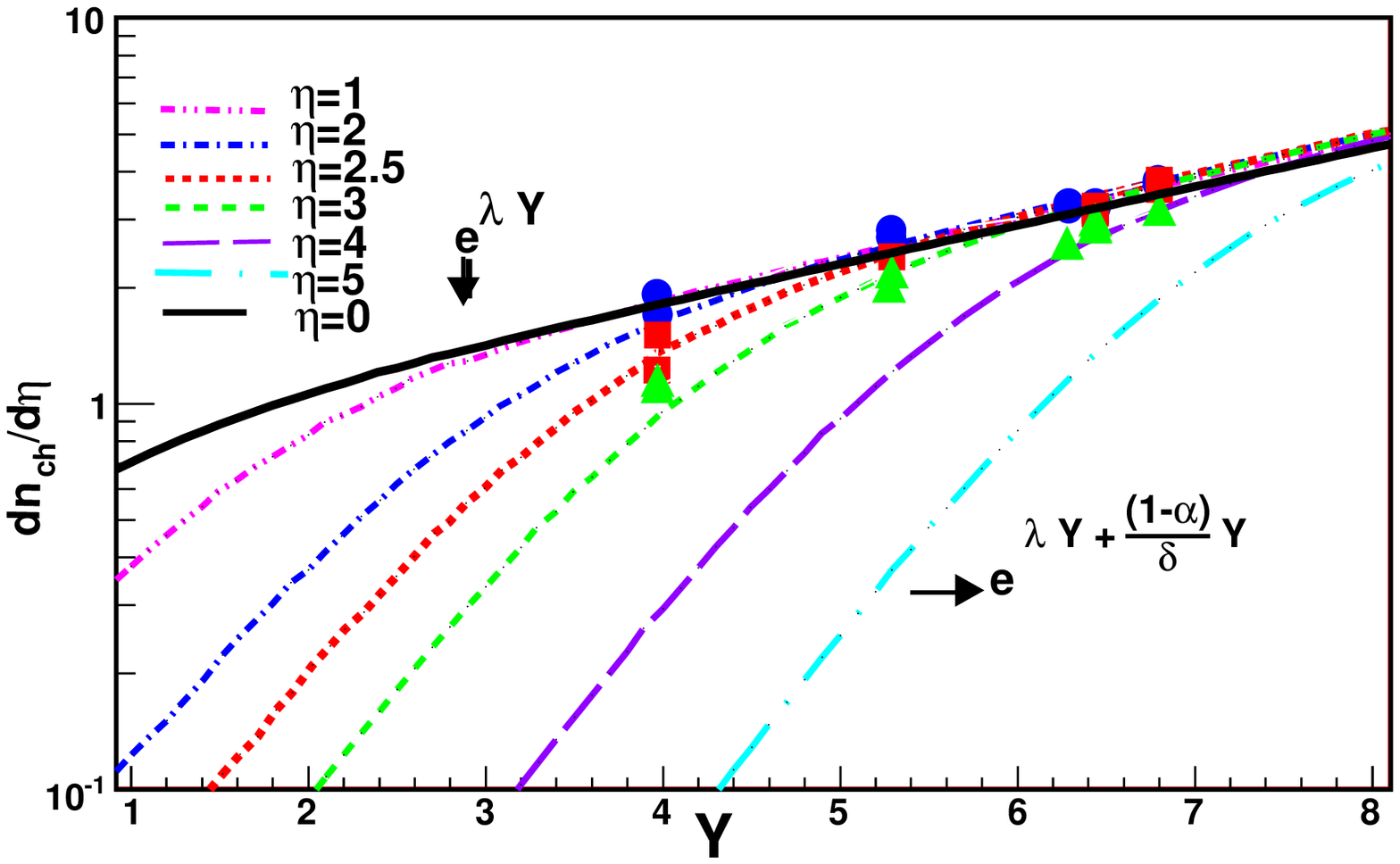}}
     \caption{Comparison of the results from the evolution of the $dn_{ch}/d\eta$ with the
     beam rapidity from equation (7) for different fixed $\eta$ values. Blue circles are
     used for data at $\eta=2$, red squares are used for data at $\eta= 2.5$, and
     green  triangles are used for data at $\eta = 3$ [15, 16].}
    \label{test4}
\end{center}
\end{figure}

Making use of (6) and (7) it is easily seen what happens. If $\eta
\geq 0$ is in the "plateau" region $dn/d\eta$ evolves as
$e^{\lambda Y}$,
\begin{equation}
\frac{dn}{d\eta}(\eta\leq (1-\alpha)Y,Y)\simeq e^{\lambda Y}.
\end{equation}
If $|\eta|$ is in the "fragmentation" region $dn/d\eta$ evolves
as $e^{\lambda Y+\frac{(1-\alpha)}{\delta}Y}$

\begin{equation}
\frac{dn}{d\eta}(\eta\geq (1-\alpha)Y,Y)\sim e^{\lambda
Y+\frac{(1-\alpha)}{\delta}Y}.
\end{equation}
For fixed $\eta$ with increasing $Y$, at some stage one falls in
the situation (9), as seen in Fig. 2 .

Our main result is that if we start at a given $Y_{0}$ with
$\frac{dn}{dy}|_{\eta>0} < \frac{dn}{dy}|_{\eta=0}$ after enough $Y$
 evolution we end up with $\frac{dn}{dy}|_{\eta>0} \simeq \frac{dn}{dy}|_{\eta=0}$
 (see Fig.2, where we have worked not with $\frac{dn}{dy}$ but with $\frac{dn}{d\eta}$).
 This means that a "plateau" proportional to $Y$ is developing.

 Such "plateau"
may be responsible for forward- backward long range
 rapidity correlations and the ridge phenomenon.
 It is well known, see for instance [17,5], that in simple models
 the forward-backward correlation parameter $b$ depends on the
 "plateau" density. If we have a long distance true plateau, then
 $b$ is constant over a long distance. On the other hand the
 height of the ridge structure or the near ridge particles is also
 dependent on particle density. The same occurs in the frame of
 Color Glass Condensate model [18].

We discuss next the question of limiting fragmentation. For $\eta$
large,
 $\eta>(1-\alpha)Y$ we have
\begin{equation}
\begin{split}
\frac{dn}{d\eta}\simeq e^{\lambda Y + \frac{(1-\alpha)}{\delta}Y - \frac{\eta}{\delta}}\\
\simeq e^{\lambda
(Y-\eta)+\frac{(1-\alpha)}{\delta}Y+\frac{(1-\lambda
\delta)}{\delta}(-\eta)}
\end{split}
\end{equation}
 The limiting fragmentation condition, i.e. the particle density being only
  a function of $Y-\eta$, is
$ 1- \alpha = 1- \lambda \delta$, or
\begin{equation}
\alpha/ \lambda= \delta.
\end{equation}

Experimentally $\alpha/ \lambda > 1$ and $\delta < 1$ and including errors,
$\alpha / \lambda = 1.17 \pm 0.15$ and $\delta = .61 \pm 0.15$, we see that
(12) is not satisfied.

To see what is expected we write (11) in the form
\begin{equation}
\frac{dn}{d\eta} \sim
e^{\lambda(Y-\eta)+\frac{(1-\alpha)}{\delta}(Y-\eta)+[\frac{(1-\lambda
\delta)}{\delta}-\frac{(1-\alpha)}{\delta}](-\eta)},
\end{equation}
and $\frac{dn}{d\eta} \rightarrow 0 $ as $ \eta $ (or $Y$) goes to
infinity.
 In this limit the distribution becomes close to a step function.
 It looks as if the fast particles in the fragmentation region
 disappear to feed up the front region in the plateau.

The arguments developed here for pp scattering, also apply to AA
and pA collisions, to the extent that they have similar
longitudinal structure.

\vspace{30mm}

\begin{large}{Acknowledgments}\end{large}

We thank J. G. Milhano for useful discussions. J. D. D. thanks the
support of the FCT/Portugal project PPCDT/FIS/575682004. I. B.
thanks the support of the FCT/Portugal SFRH/BD/51370/2011. I. B,
and C. P. were supported by the project FPA 2008-01177 and FPA 2011-22776 of MICINN,
the Spanish Consolider Ingenio 2010 program CPAN and Conselleria
Educacion Xunta de Galicia.

\vspace{10mm}
\begin{Large}{References}\end{Large}

\end{document}